\pdfoutput=1
\documentclass[3p,times,twocolumn]{elsarticle}

\usepackage{ecrc}
\usepackage{subfigure}

\volume{00}

\firstpage{1}

\journalname{Nuclear Physics B Proceedings Supplement}

\runauth{X.~Feng, K.~Jansen, M.~Petschlies and D.~Renner}

\jid{nuphbp}

\jnltitlelogo{Nuclear Physics B Proceedings Supplement}

\usepackage{amssymb}
\usepackage[figuresright]{rotating}

\begin{document}

\begin{frontmatter}



\dochead{}

%\title{}

\title{Hadronic Vacuum Polarization Contribution to g-2 from the Lattice
\vspace*{-1.9cm}
\begin{flushright}
{\small DESY 11-236, JLAB-THY-12-1411, KEK-CP-264} 
\end{flushright}
\vspace*{1.9cm}}

\author[label1]{X. Feng}
\author[label2]{K. Jansen}
\author[label3]{M. Petschlies}
\author[label4]{D. Renner}
\address[label1]{High Energy Accelerator Research Organization (KEK), Tsukuba 305-0801, Japan}
\address[label2]{NIC, DESY, PLatanenallee 6, 15738 Zeuthen, Germany}
\address[label3]{The Cyprus Institute, P.O.Box 27456, 1645 Nicosia, Cyprus}
\address[label4]{Jefferson Lab, 12000 Jefferson Avenue, Newport News, VA 23606, USA}

%\author{}

%\address[\label1]{}

\begin{abstract}
We give a short description of the present situation 
of lattice QCD simulations.  
We then focus on the computation of the anomalous magnetic moment 
of the muon using lattice techniques. 
We demonstrate that by employing improved observables for 
the muon anomalous magnetic moment, a significant reduction 
of the lattice error can be obtained. This provides a promising 
scenario that the accuracy of lattice calculations can match 
the experimental errors.
\end{abstract}


\end{frontmatter}

% \linenumbers

\section{introduction}
\label{introduction}
The interaction between quarks at large distances becomes strong 
such that 
analytical methods as perturbation theory fail to analyze QCD.
A method to nevertheless tackle the problem is to formulate QCD on
a 4-dimensional, euclidean space-time grid. This setup first of all
allows for a rigorous definition of QCD and leads to fundamental
theoretical and conceptual investigations. On the other hand, the lattice approach
enables theorists to perform large scale numerical simulations.

In the past, lattice physicists had to work with a number of limitations
when performing numerical simulations which turn out to be 
extremely expensive, leading to the need for Petaflop computing and
even beyond, a regime
of computing power we just reach today.
Therefore, for a long time the sea quarks were treated as infinitely heavy,
indeed a crude approximation given that the up and down quarks have
masses of only O(MeV). In a next step, only the lightest quark doublet,
the up and down quarks, were taken into consideration, although their mass values
as used in the simulation had been unphysically large.

\begin{figure}[t]
  \centering
  {\includegraphics[width=1.00\linewidth]{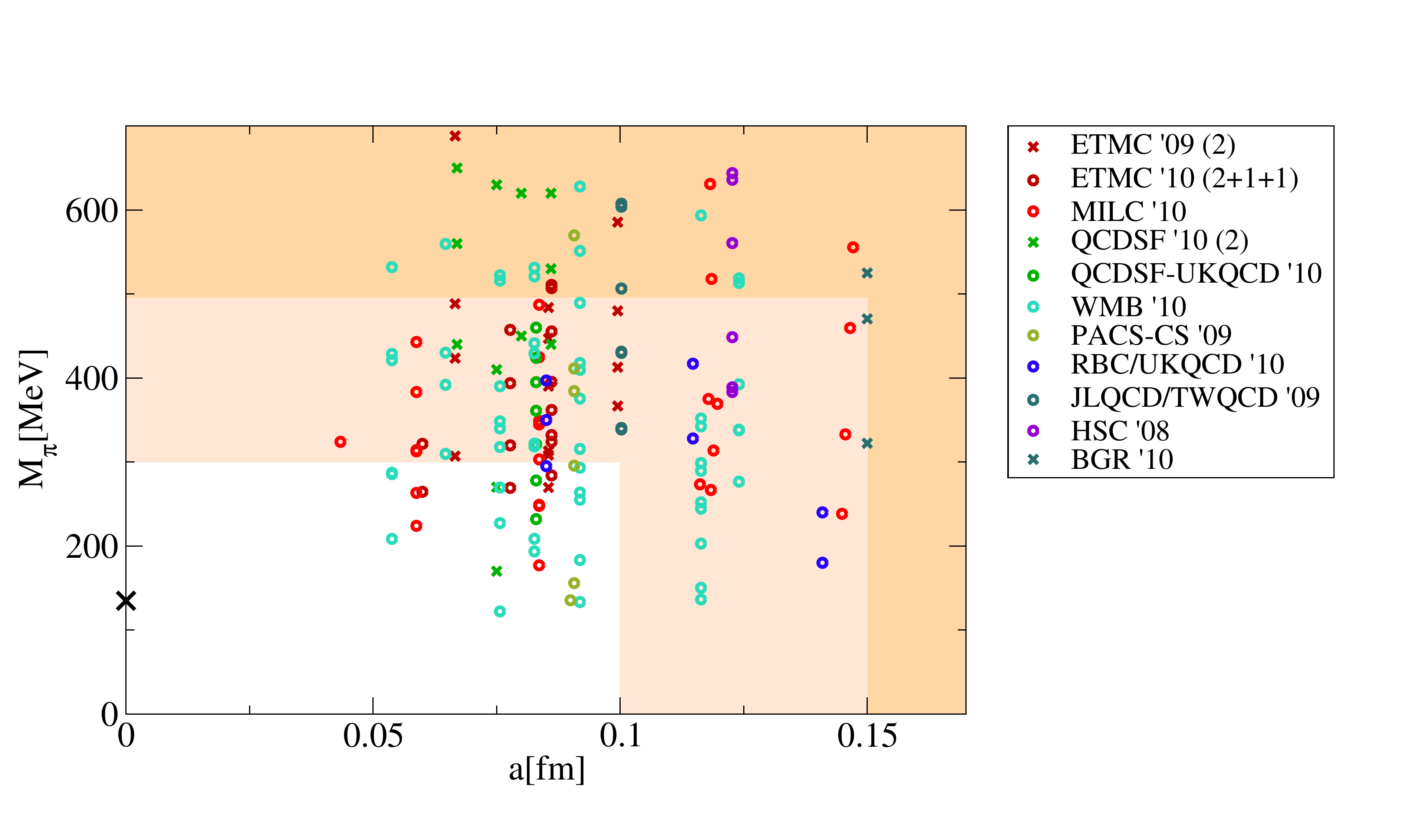}}
  \caption{
The values of the lattice spacing $a$ and pion masses
$m_\pi$ as
employed presently in typical QCD simulations by various collaborations 
as listed in the legend. The cross denotes the physical point. 
The figure is taken from \cite{Hoelbling:2011kk} where also further details and references
to the various collaborations can be found}. 
  \label{fig:landscape}
\end{figure}

Nowadays, besides the up and down quarks, also the strange quark is
included in the simulations. In addition, these simulations are performed
in almost physical conditions, having the quark masses close to their
physical values, large lattices with about 3fm linear extent and
small values of the lattice spacing such that a controlled continuum limit
can be performed. The situation of present days simulation
landscape is illustrated in fig.~\ref{fig:landscape}, taken from \cite{Hoelbling:2011kk}. 
In the figure, the black cross indicates the physical point. 

The fact that presently simulations close to the physical 
situation can be performed is due to three
main developments:
$i)$ algorithmic breakthroughs which gave a substantial factor 
$O(> 10)$ of improvement;
$ii)$ machine
development with a computing power of the present BG/P systems which is even outperforming
Moore's law, $iii)$ conceptual developments, such as the use of
improved actions which reduce lattice artefacts
and the development of non-perturbative renormalization.

\begin{figure}[t]
  \centering
  {\includegraphics[width=0.45\linewidth]{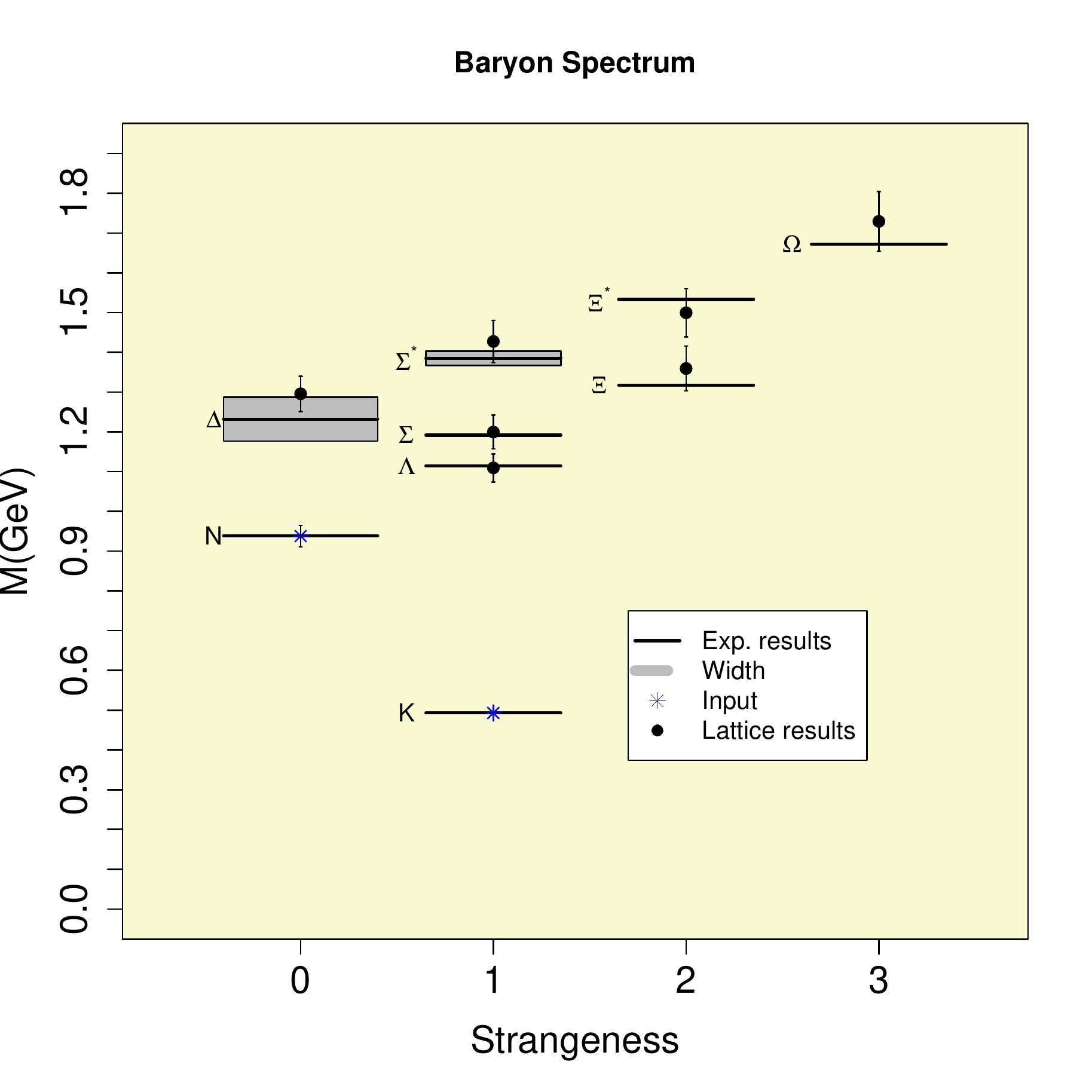}}
  \caption{
The continuum strange baryon
spectrum from the ETM collaboration \cite{Alexandrou:2009qu} using $N_f=2$
flavours of quarks with only mass-degenerate up and down quark masses.}
  \label{fig:spectrum}
\end{figure}

As an example of physical results we can achieve presently,
we show in fig.~\ref{fig:spectrum} the continuum extrapolated strange baryon spectrum
as obtained by the 
European Twisted Mass
Collaboration (ETMC) \cite{Alexandrou:2009qu} of which the authors
are members. 
The first complete calculation of the baryon spectrum was achieved
by the BMW collaboration \cite{Durr:2008zz} and nowadays a number of lattice 
groups
are providing calculations of the hadron masses and nucleon structure, 
see e.g. \cite{Alexandrou:2011iu} for a recent review. 
The baryon spectrum
calculation has been considered a benchmark study for lattice QCD
for a long time.
It is therefore very reassuring that finally this important result can be
obtained precisely from ab-initio and non-perturbative lattice simulations.

\section{The anomalous magnetic moment of the muon}

The progress of lattice calculations discussed in the previous 
chapter motivates to address more demanding quantities 
than the baryon masses. A prime example is 
the anomalous magnetic moment of the 
muon $a_\mu\equiv (g_\mu-2)/2$. The reason for looking at this quantity is that here  
theory and experiment disagree and 
one finds
$a_\mu^{\rm exp}-a_\mu^{\rm th} = 2.90 (91)\times 10^{-9}$
which leads to a larger than $3\sigma$ level discrepancy.

Clearly, this
is a very interesting result. It means that either
in the theoretical calculation something has been neglected or
has not properly been included. Or, somewhat much more exciting,
the discrepancy points to a breakdown of the standard model of
particle interactions and the inconsistency stems from effects
of some yet unknown new physics beyond the standard model.

Indeed, calculations show that these new physics effects
would lead to a correction to the anomalous magnetic
moment of size

\begin{equation}
  \label{eq:amunewphysics}
    \delta(a_l^{\rm new physics}) = m_{\rm lepton}^2/M_{\rm new physics}^2\; .
\end{equation}
Here $m_{\rm lepton}$ is the mass of one of the leptons
and $M_{\rm new physics}$ represents the mass (or scale)
of a particle originating from the (unknown)
new physics beyond the standard model.
The formula in eq.~(\ref{eq:amunewphysics}) shows that in the case of the
muon anomalous magnetic moment the effect of new physics
would show up about $(m_\mu/m_e)^2\approx 4\cdot 10^4$ times
stronger than in the case
of the electron. In principle, the $\tau$-lepton would be even
more suitable to detect these new physics effects, but
unfortunately due to the very short lifetime
of the $\tau$ lepton the experimental measurements of
the anomalous magnetic moment of the $\tau$
are presently much too imprecise to unveil a possible new physics
contribution.
This leaves us then with the muon anomalous magnetic moment as the
ideal place to look for new physics and indeed a large number of works
has been devoted to explore this possibility, see \cite{Jegerlehner:2009ry}.

\section{When the lattice enters the game}

It has been found that electromagnetic and weak interaction 
effects can by far not serve as an explanation of this 
discrepancy \cite{Jegerlehner:2009ry}. 
However, the strong interaction can have a large effect
since the hadronic
contributions $a_\mu^{\rm had}$ dominate the uncertainty of 
the standard model value.
The problem is that the strong interactions of quarks and gluons in QCD  
are intrinsically
of non-perturbative nature. Taking these contributions into account
by perturbation theory is therefore rather doubtful.
Employing additional model assumptions to estimate
$a_\mu^{\rm had}$ will not provide a
fully controlled and reliable calculation of the hadronic
contributions and hence an unambiguous and stringent test whether the standard model
is correct or must be extended by some new physics cannot be performed.

It is exactly at this point where lattice field theory methods
applied to quantum chromodynamics can help -- at least in principle.
In lattice QCD the theory is formulated on a discrete
4-dimensional euclidean space-time lattice and the theory
is approached by means of numerical simulations.
It goes beyond the scope of this article to explain the mathematical
concepts of lattice QCD but such numerical simulations allow then
to compute physical quantities in a fully non-perturbative fashion
without relying on any model assumptions or approximations.

Of course, the discretization itself induces a systematic error
which must be removed by making the lattices finer and finer until
the continuum of space time points is recovered by some suitable
extrapolation process, a procedure which
is called the {\em continuum limit}.
In addition, the simulations necessarily demand a finite number
of lattice points which can lead to {\em finite size effects} when
the lattice is not large enough in physical units. Finally,
often the simulations need to be performed at values of hadron
masses that are larger than the ones observed in nature. The reason is
that for smaller and smaller hadron masses the computational costs increase
rapidly such that one is restricted to values of, say,
pion masses that are a factor of about two larger than the
ones observed in nature.

\begin{figure}[t]
  \centering
  {\includegraphics[width=0.60\linewidth]{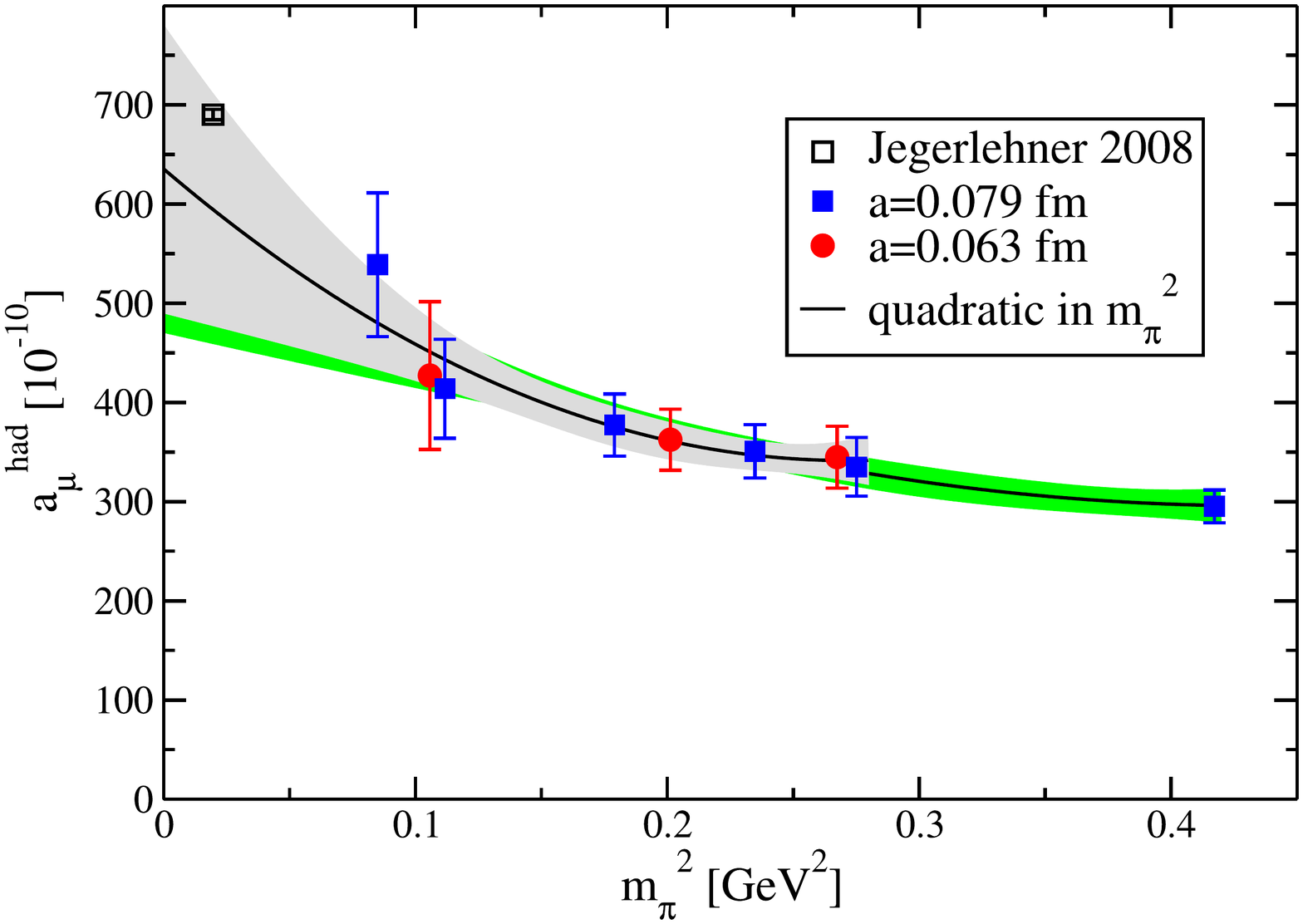}}\quad
  \caption{We show an earlier result for the hadronic contribution
    to the muon anomalous magnetic moment
    computed by the authors.} 
  \label{fig:amulattold}
\end{figure}

All these systematic effects that appear in lattice simulations
need to be controlled in a quantitative way. For example,
to reach physical values of the pion masses, an extrapolation
to the physical point where the pion mass assumes its
physical value needs to be performed. This appeared to be very
problematic in the past. This is illustrated in
fig.~\ref{fig:amulattold}. The figure shows that the lattice
results for $a_\mu^{\rm had}$ are significantly below
the experimental number. An extrapolation to the physical point,
reconciling the lattice data with experiment becomes in this
situation very difficult and even needs some additional model
assumptions. This all leads to an error of $a_\mu^{\rm had}$ as
obtained from lattice simulations that is about a factor of 10 larger
than the phenomenological one \cite{Renner:2009by}. The lattice community
have been therefore rather sceptical in the past that lattice QCD can
provide a significant contribution to our understanding of the
discrepancy in $g_\mu-2$.

One additional suspicion has been that so-called dis-connected (singlet)
contributions could be substantial. In all existing lattice
calculations these contributions were neglected, however.
The reason is simply that these contributions are very noisy and
therefore hard to compute reliably.
Nevertheless, in ref.~\cite{Feng:2011zk} a dedicated effort
has been undertaken 
to calculate for the first time these contributions.
As a result, it could be established that the dis-connected
contributions are in fact small and can be
safely neglected.
In addition, also the effects of non-zero values of the lattice spacing
and the finite volume turned out to be small.
Thus the difficulty to reconcile lattice data
with the experimental result, shown in fig.~\ref{fig:amulattold},
is rather puzzling.

A resolution of this puzzle was only given this year 
by us \cite{Feng:2011zk}. We observed that by a suitable redefinition
of the lattice observable needed to compute $a_\mu^{\rm had}$ a much
smoother and much better controlled approach to the physical point can be
achieved.

\begin{figure}[t]
  \centering
  {\includegraphics[width=0.60\linewidth]{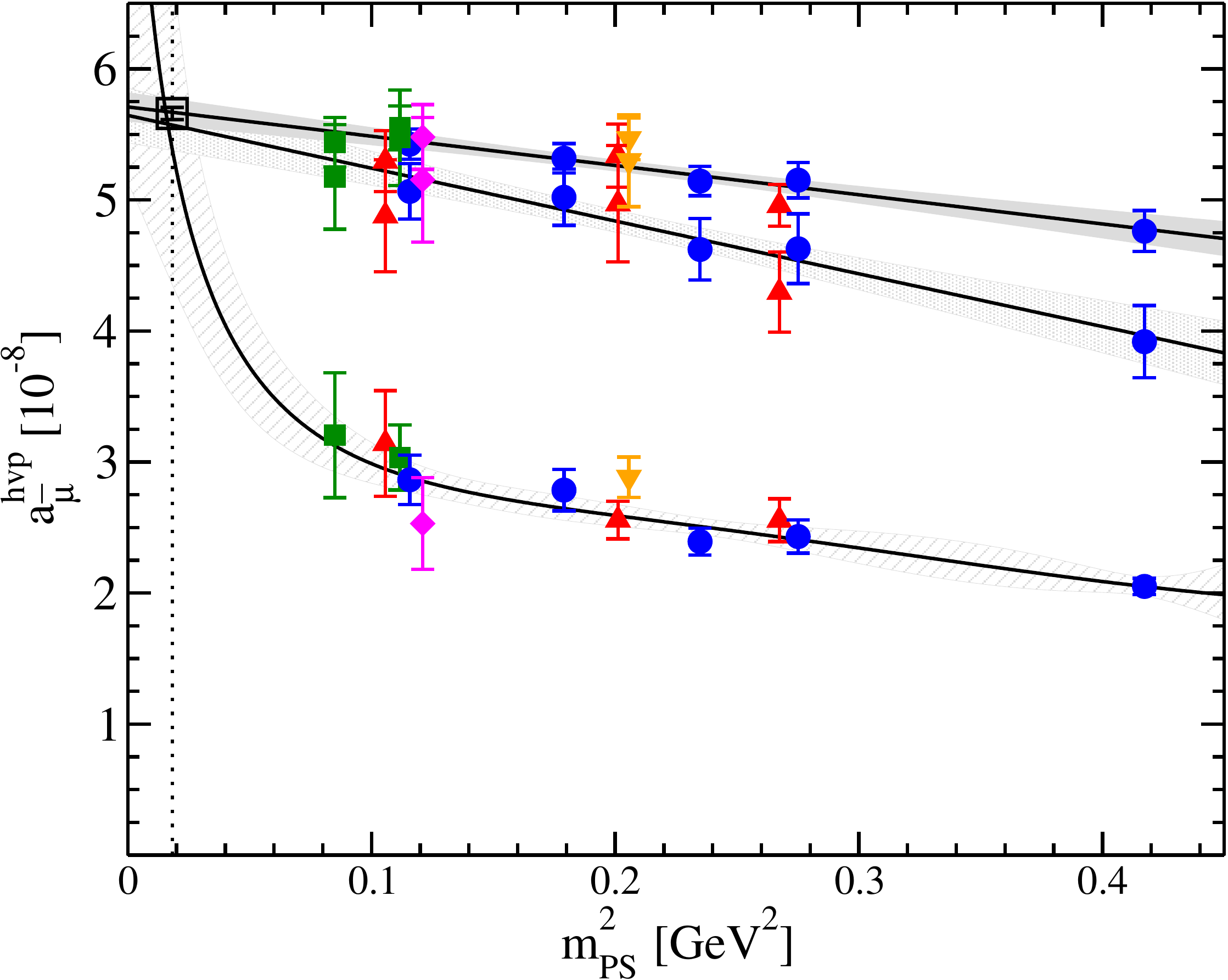}}
  \caption{
    The present day status of computing 
    $a_\mu^{\rm had}$ is represented using the
    improved observables discussed in the text. 
    The curves correspond to the definitions $r_1$,
    $r_2$ and $r_3$ in eqs.~(\ref{eq:rchoices}) from bottom to top.
}
  \label{fig:amulattnew}
\end{figure}

To illustrate the idea, let us give the definition of
$a_\mu^{\rm had}$,

\begin{equation}
a_\mu^{\rm had} = \alpha^2\int_0^\infty dQ^2 \frac{1}{Q^2} \omega(r)\Pi_R(Q^2)\; .
\label{eq:amudef}
\end{equation}
Here $\alpha$ is the electromagnetic coupling and $\Pi_R(Q^2)$ the renormalized
vacuum polarization function,
$\Pi_R(Q^2) = \Pi(Q^2)-\Pi(0)$. The functional form of $\omega(r)$ is
analytically known and the argument $r$ is given by $r=Q^2/m_\mu^2$
where $m_\mu$ denotes the mass of the muon and $Q$ a generic momentum.
The key observation is now that on the lattice there is a large freedom
to choose a definition of $r$. The only requirement is that in the limit
of reaching a physical pion mass the continuum definition of
$r=Q^2/m_\mu^2$ is recovered. Hence, one may define

\begin{equation}
r_{\rm latt} = Q^2\cdot \frac{H^{\rm phys}}{H}
\label{eq:rdef}
\end{equation}

with possible choices for $H$

\begin{eqnarray}
r_1: & H=1 & H^{\rm phys}=1/m_\mu^2 \nonumber \\
r_2: & H=m_V^2(m_{\rm PS}) & H^{\rm phys}=m_\rho^2/m_\mu^2 \nonumber \\
r_3: & H=f_V^2(m_{\rm PS}) & H^{\rm phys}=f_\rho^2/m_\mu^2\; .
\label{eq:rchoices}
\end{eqnarray}
Here, $m_V(m_{\rm PS})$ is the mass of the $\rho$-meson and $f_V(m_{\rm PS})$ the
$\rho$-meson decay constant as determined on the lattice at unphysical
pion masses $m_{\rm PS}$. Furthermore, $m_\rho$ and $f_\rho$ denote
the corresponding $\rho$-meson mass and decay constant at the physical point.
All the definitions in eqs.~(\ref{eq:rchoices})
guarantee that indeed the desired definition
of $r$ is recovered in the limit of a physical pion mass since then
by definition $m_V(m_{\rm PS})$ and $f_V(m_{\rm PS})$
assume their physical values.
In fig.~\ref{fig:amulattnew} we show the results
for $a_\mu^{\rm had}$ for all three definitions of $r$. Clearly,
for the definitions $r_2$ and $r_3$ the behaviour of the lattice data
towards physical pion masses is simply linear and allows for a controlled
extrapolation to the physical point. As a result, one finds using definition
$r_2$ in eq.~(\ref{eq:rchoices}) values from the lattice computations and
experiment
\begin{eqnarray}
a_{\mu,N_f=2}^\mathrm{had,latt} & =  & 5.72\,(16)\cdot 10^{-8} \nonumber \\
a_{\mu,N_f=2}^\mathrm{had,exp} & = & 5.66\,(5)\cdot 10^{-8}\; .
\label{eq:lattcompare}
\end{eqnarray}
In the equations above, the index $N_f=2$ indicates that
in the lattice QCD calculations only a mass-degenerate pair of up
and down quarks were used. Since the strange and charm quark
flavours were neglected the simulations do not correspond to a fully
physical situation which also leads to some ambiguity 
in the experimental extraction of $a_{\mu,N_f=2}^\mathrm{had,exp}$. 
This shortcoming needs to be overcome
in the future.

\section{Conclusion} 

In conclusion, using the modified and improved definitions of
$a_\mu^{\rm had}$ on the lattice it is not only possible
to recover the experimental result. As the comparison in eq.~(\ref{eq:lattcompare})
shows it is
now also possible to come significantly closer to the experimental accuracy.
The idea of the improved observables which led to a much reduced
error for an important quantity such as $g_\mu -2$
has therefore led to the promising situation that lattice QCD 
calculations can match the experimentally obtained errors. 

As already mentioned above, the results
for $a_\mu^{\rm had}$ here discussed have been achieved for the
case of two mass-degenerate quark flavours. What is needed in the future
is the inclusion of the strange and the charm quarks to allow for a direct
comparison to the experimental results. In addition, newly planned experiments
at Fermilab \cite{Roberts:2004mv} and
J-PARC \cite{Sato:2011zz} are aiming at an accuracy of below $0.5$\%
for the hadronic contribution to the muon anomalous magnetic moment. 
To match
this accuracy dedicated lattice simulations have to be performed 
on large enough volumes and as close
as possible to the physical point. 
In addition, explicit
effects of isospin breaking and electromagnetism might need to be 
included. All this is in principle
reachable within lattice QCD but, it constitutes a real challenge for
the lattice community. An even larger challenge are contributions
to $g_\mu -2$ that appear at higher order of the electromagnetic
coupling, most notably the so-called light-by-light contributions.

However, a number of lattice groups 
\cite{Feng:2011zk,Boyle:2011hu,Mainznew} are working on 
this problem presently and it can be expected that the lattice 
will provide a significant contribution to answer the question, 
whether the observed discrepancy in the muon anomalous magnetic moment 
is indeed a sign of new physics. 

\section*{Acknowledgments}
This work is coauthored in part by Jefferson Science Associates, LLC under U.S.
DOE Contract No. DE-AC05-06OR23177.



%% References

%\nocite{*}
\bibliographystyle{elsarticle-num}
\bibliography{paper}






% \end{thebibliography}

\end{document}